\documentclass[final,12pt,authoryear]{elsarticle}
\usepackage{amssymb}
\usepackage{amsmath}
\usepackage{amsmath, amssymb}
\usepackage{natbib}
\usepackage{hyperref}
\makeatletter
\def\ps@pprintTitle{%
  \let\@oddhead\@empty
  \let\@evenhead\@empty
  \def\@oddfoot{}%
  \let\@evenfoot\@oddfoot}
\makeatother
\journal{}
\usepackage{geometry}

\begin{document}

\begin{frontmatter}

\title{Single-Parameter Exponential Model Outperforms Standard Dark Matter and Modified Newtonian Frameworks in Predicting Galactic Rotation Curves\tnoteref{t1}}

\tnotetext[t1]{Received 18 October 2025; Accepted 19 December 2025; Published 23 December 2025.\\
This version of the article has been accepted for publication, after peer review, and is subject to Springer Nature's AM terms of use, but is not the Version of Record and does not reflect post-acceptance improvements, or any corrections. The Version of Record is published in \textit{Astrophysics and Space Science} (2025) and is available online at: \url{https://doi.org/10.1007/s10509-025-04531-7}}

\author{Mohammed Guhdar}

\affiliation{organization={Department of Computer Science, University of Zakho},
            country={Kurdistan Region, Iraq}}

\begin{abstract}
The functional forms of galactic rotation curves provide a fundamental test for models of gravity and dark matter. In this work, we conduct a rigorous statistical comparison of five distinct models against the complete sample of 175 galaxies from the SPARC database. The models under consideration are: (1) a one-parameter empirical law, physically constrained by each galaxy's measured flat disc rotation velocity ($V_{\text{flat}}$); (2) a classical two-parameter Freeman disk; (3) a three-parameter Navarro-Frenk-White (NFW) dark matter halo plus baryons model; (4) a one-parameter Modified Newtonian Dynamics (MOND) model; and (5) a two-parameter Bosma/Dark Matter Disk model. We employ a Bayesian framework with Markov Chain Monte Carlo (MCMC) for parameter estimation, and use the Akaike Information Criterion (AICc) to assess the goodness-of-fit and statistical parsimony of the best-fit models. Our results reveal a decisive preference for the constrained one-parameter empirical model. It is selected as the best descriptor for a clear majority of the sample (60.6\%) and, critically, also achieves the best median goodness-of-fit, with a reduced chi-squared of $\chi^2_{\nu} = 0.41$. In contrast, the standard NFW framework is preferred in only 12.6\% of cases, while the MOND and Freeman disk models are selected as optimal for just 10.3\% and 9.1\%, respectively. The Bosma model, despite linking dynamics to the gas distribution, is preferred in only 7.4\% of cases. The pronounced statistical and descriptive success of this simple, physically-anchored law over more complex, established theories suggests that its functional form represents a fundamental and highly efficient organizing principle in the dynamics of disk galaxies.
\end{abstract}


\end{frontmatter}

\section{Introduction}
\label{sec:introduction}

Elucidating the formation, structure, and evolution of galaxies remains a cornerstone of modern astrophysics and cosmology. Within the prevailing $\Lambda$ Cold Dark Matter ($\Lambda$CDM) paradigm, galaxies are believed to form hierarchically within extended halos composed primarily of non-baryonic dark matter \citep{White1978}. The gravitational potential wells established by these halos are considered essential for trapping baryonic gas, allowing it to cool, condense, and form the stars and galactic structures we observe today.

An early and foundational model for the visible component of disk galaxies is the exponential disk, often referred to as the Freeman disk \citep{freeman1970disks}. This model, characterized by an exponentially declining surface density, predicts a rotation curve that rises to a peak and then typically declines at larger radii. While it provides a good description for the stellar component, its velocity profile alone is insufficient to explain the observed flatness of rotation curves in their outer regions, a key observation that has historically pointed towards the need for additional unseen mass or modified gravitational laws \citep{binney2008galactic}.

One of the most compelling pieces of evidence underpinning the need for dark matter, or alternatively, a modification of gravitational laws, arises from the study of galaxy rotation curves – the measurement of orbital velocities of gas and stars as a function of galactocentric radius, $r$. Pioneering work by astronomers Vera Rubin and Albert Bosma in the 1970s and 1980s \citep{Bertone2018} revealed a significant discrepancy: the rotation velocities in the outer regions of spiral galaxies remain remarkably flat or even slightly rising, extending far beyond the radii where the gravitational influence of the visible baryonic matter (stars and gas) should cause a Keplerian decline ($v \propto r^{-1/2}$) \citep{Rubin1980, Bosma1981}. This kinematic anomaly strongly indicates the presence of substantial unseen mass or a deviation from Newtonian dynamics on galactic scales.

The standard approach within $\Lambda$CDM posits that this discrepancy is resolved by the gravitational influence of the dark matter halo. Cosmological N-body simulations predict that these halos should exhibit near-universal density profiles, often approximated by the Navarro-Frenk-White (NFW) profile \citep{Navarro1996, Navarro1997} or similar forms (e.g., Einasto profile) \citep{ludlow2016einasto}. While $\Lambda$CDM has achieved remarkable success in explaining large-scale structure formation, the cosmic microwave background anisotropies, and the overall galaxy population statistics, persistent challenges remain on galactic scales. These include the "cusp-core" problem (simulated halos often predict central density cusps steeper than sometimes observed) \citep{del2021review}, the "missing satellites" problem, and the "too big to fail" problem, among others, suggesting potential tensions or incompleteness in the standard model or our understanding of baryonic feedback processes \citep{Bullock2017}.

An alternative framework, Modified Newtonian Dynamics (MOND), postulates a modification to Newtonian gravity or inertia below a characteristic acceleration scale $a_0 \approx 1.2 \times 10^{-10}$ m s$^{-2}$ \citep{Milgrom1983a}. In the deep MOND regime ($a \ll a_0$), the effective gravitational force scales differently, naturally leading to asymptotically flat rotation curves without invoking dark matter. MOND has demonstrated considerable phenomenological success in explaining various galactic scaling relations, such as the Baryonic Tully-Fisher Relation, and often provides excellent fits to individual galaxy rotation curves with minimal parameter tuning \citep{Famaey2012, McGaugh2014}. However, MOND faces its own challenges, particularly in explaining the dynamics of galaxy clusters and integrating seamlessly into a comprehensive cosmological model \citep{Bekenstein2004}.

In light of these ongoing discussions, the exploration of alternative descriptions for galaxy rotation curves remains a valuable endeavor. Such data-driven discoveries serve as powerful, unbiased benchmarks against which more complex theories can be tested. Recently, for example, \citep{taylor2025empirical} demonstrated that a multiplicative polynomial correction factor can also systematically improve the fits of canonical halo models by using four free parameters, further highlighting the presence of universal, unmodeled systematics in the observational data,  Through extensive empirical testing of various functional forms against galaxy rotation curve data, we identified a remarkably simple and effective one-parameter law. This work proposes a direct investigation into the descriptive and predictive power of this law, which takes the form:
\begin{equation}
    v(r) = V_{\text{flat}} \left( 1 - \exp\left(-\frac{r}{r_{\text{scale}}}\right) \right)
    \label{eq:main_equation}
\end{equation}
Here, $v(r)$ is the rotational velocity at a given radius $r$. The model's power lies in its structure: the characteristic asymptotic velocity, $V_{\text{flat}}$, is fixed to a value derived from the kinematic data of each galaxy. This leaves $r_{\text{scale}}$, a core radius defining the transition scale, as the \textit{sole free parameter} to be determined by the fit.

The primary objective of this work is to quantify the power of Equation~\ref{eq:main_equation} by applying it to the extensive, high-quality SPARC database \citep{Lelli2016}. A central component of our analysis is a rigorous, quantitative comparison of this one-parameter empirical law against established theoretical benchmarks: a standard NFW dark matter halo, a MOND formulation, and a classical Freeman disk model. To ensure a fair comparison, the baryonic mass-to-light ratios for the NFW and MOND models are treated as free parameters within our Bayesian fitting framework.

Notwithstanding the general success of any given model, a comprehensive analysis requires a systematic examination of its limitations and failures. An important facet of this investigation is the identification of galaxies whose kinematics are poorly described by the proposed law and by the established theoretical models. A detailed analysis reveals that the most pronounced failures across all models consistently occur for galaxies with high baryonic mass \citep{read2005baryonic} and large rotational velocities ($V_{\text{flat}} \gtrsim 200$ km/s), such as UGC02487, UGC09133, and NGC6674. This systematic investigation of model failures serves to delineate the boundaries of their applicability and to highlight specific astrophysical regimes that demand more sophisticated theoretical treatment.

The overarching aim is thus to demonstrate the remarkable statistical efficiency and accuracy of the proposed one-parameter law. By showing its frequent superior performance over more complex and physically-specific theories, we establish its value as a powerful, data-derived benchmark and motivate a deeper inquiry into the physical origins of its success.

\subsection*{Principal Contributions}
The principal contributions of this investigation are delineated as follows:
\begin{itemize}
    \item The validation of a concise, one-parameter empirical law (Eq.~\ref{eq:main_equation}) for describing galactic rotation curves, conducted within a robust Bayesian MCMC framework.
    \item A comprehensive demonstration of this law's high fidelity in reproducing observed kinematics, including its superior statistical preference (as determined by the AICc) when compared against standard NFW, MOND, Bosma/Dark Matter Disk and Freeman disk models.
    \item The systematic quantification of the single best-fit parameter, $r_{\text{scale}}$, across the sample, providing a fundamental characterization of the kinematic transition scale in disk galaxies.
    \item A direct statistical confirmation of the law's notable descriptive efficiency and accuracy, highlighting its ability to capture complex kinematics with minimal parametric freedom.
    \item An identification and analysis of galaxies that prove challenging for the proposed law and its theoretical competitors, thereby highlighting specific astrophysical regimes where current models are insufficient.
    \item The establishment of this simple, data-derived equation as a powerful new benchmark for testing more elaborate astrophysical theories and motivating future work to uncover the physical basis for its effectiveness.
\end{itemize}

\section{Data and Methodology}
\label{sec:data_method}

\subsection{Observational Data: The SPARC Sample}
\label{subsec:data}

To ensure a rigorous evaluation of the competing models, we utilize high-precision observational data from the Spitzer Photometry and Accurate Rotation Curves (SPARC) database \cite{Lelli2016}. This catalog provides a unique combination of extended H\textsc{i} rotation curves and near-infrared surface photometry, allowing for a precise decomposition of baryonic mass components.

We analyzed the complete SPARC catalog of 175 galaxies. By retaining the full sample rather than applying restrictive pre-selection cuts, we ensure that our comparative analysis is free from selection bias and representative of the widest possible range of galactic morphologies, from high-surface-brightness spirals to low-mass irregulars. While previous studies often exclude galaxies with lower quality scores, our Bayesian framework is sufficiently robust to handle varying data quality by incorporating observational uncertainties directly into the likelihood function. The characteristic asymptotic velocity, $V_{\text{flat}}$, used to physically constrain our primary model, was derived directly from the data for each galaxy by calculating the median velocity of its outer stable region.

\subsection{Bayesian Modeling Framework}
\label{subsec:fitting}

The core of our methodology is a full Bayesian parameter inference for five distinct models. This approach allows us to not only find the most probable parameters but also to rigorously quantify their uncertainties. The analysis employs a Markov Chain Monte Carlo (MCMC) framework \citep{edwards2010markov} to explore the parameter space of each model ($\mathbf{p}$) and map out the posterior probability distribution, $P(\mathbf{p} | \text{data}) \propto \mathcal{L}(\text{data} | \mathbf{p}) P(\mathbf{p})$, where $\mathcal{L}$ is the likelihood and $P(\mathbf{p})$ is the prior.

The sampling is guided by the log-probability, $\ln P(\mathbf{p} | \text{data}) = \ln \mathcal{L}(\text{data} | \mathbf{p}) + \ln P(\mathbf{p})$. Assuming Gaussian errors, the log-likelihood function is proportional to the standard chi-squared statistic, $\chi^2$:
\begin{equation}
    \ln \mathcal{L}(\mathbf{p}) = -\frac{1}{2} \chi^2 = -\frac{1}{2} \sum_{i=1}^{N} \left[ \frac{v_{\text{obs},i} - v_{\text{model}}(r_i; \mathbf{p})}{\sigma_i} \right]^2
    \label{eq:chi_squared}
\end{equation}
Here, $v_{\text{model}}(r_i; \mathbf{p})$ is the velocity predicted by the model at radius $r_i$ given the parameters $\mathbf{p}$. The total uncertainty, $\sigma_i$, is calculated by adding a constant systematic error floor, $\sigma_{\text{floor}} = 5.0 \, \text{km s}^{-1}$, in quadrature to the published observational error, $\delta v_{\text{obs},i}$, such that $\sigma_i^2 = \delta v_{\text{obs},i}^2 + \sigma_{\text{floor}}^2$.

\subsection{Model Definitions and Priors}
\label{subsec:model_definitions}

We adopt uninformative, uniform (flat) priors for all model parameters, constraining them only to physically plausible ranges. The five models are detailed below.

\begin{enumerate}
    \item \textbf{Constrained Empirical Law (Empirical, $k=1$):} 
    This is the primary model under investigation, given by $v(r) = V_{\text{flat}}(1 - \exp(-r/r_{\text{scale}}))$. The asymptotic velocity, $V_{\text{flat}}$, is fixed for each galaxy to the value derived from its kinematic data. This leaves only a single free parameter: the transition scale radius, $\mathbf{p}_{\text{Empirical}} = (r_{\text{scale}})$.

    \item \textbf{Freeman Disk Model ($k=2$):}
    This model represents the velocity profile of an isolated, infinitesimally thin exponential disk. Its velocity is given by:
    \begin{equation}
        v^2(r) = \frac{2 G M_{\text{disk}}}{r_{\text{disk}}} y^2 [I_0(y)K_0(y) - I_1(y)K_1(y)], \quad \text{where } y = \frac{r}{2 r_{\text{disk}}}
    \end{equation}
    The free parameters are the logarithm of the total disk mass and its scale length: $\mathbf{p}_{\text{FreeD}} = (\log_{10} M_{\text{disk}}, r_{\text{disk}})$.

    \item \textbf{NFW + Baryons Model ($k=3$):} 
    This model combines an NFW dark matter halo with the observed baryonic components, such that $v_{\text{model}}^2 = v_{\text{baryon}}^2 + v_{\text{NFW}}^2$. The stellar mass-to-light ratio for the disk ($\Upsilon_{\text{disk}}$) is a free parameter. The full set of free parameters is: $\mathbf{p}_{\text{NFW+B}} = (\log_{10} M_{200}, c_{200}, \log_{10} \Upsilon_{\text{disk}})$.

    \item \textbf{MOND Model ($k=1$):} 
    This model employs the "Simple" interpolation function \cite{famaey2005modified}, $\mu(x) = x/(1+x)$. The true acceleration ($g$) is related to the Newtonian acceleration ($g_N = v_{\text{baryon}}^2/r$) via the fundamental relation:
    \begin{equation}
        g \cdot \mu\left(\frac{g}{a_0}\right) = g_N
        \label{eq:mond_relation}
    \end{equation}
    This relation yields a quadratic equation for $g$, which is solved algebraically. The stellar disk mass-to-light ratio is the sole free parameter: $\mathbf{p}_{\text{MOND}} = (\log_{10} \Upsilon_{\text{disk}})$.
  \item \textbf{Generalized Bosma/Dark Matter Disk Model ($k=2$):}
    To test the hypothesis that galaxy dynamics are driven by the baryonic mass distribution in the disk \cite{SylosLabini2024, Bosma1981}, we adopt a generalized formulation of the Bosma effect. While classical approaches often fix the gas contribution, we treat both the stellar and gas mass-to-light ratios as free parameters to allow for maximal model flexibility (e.g., accounting for scaling due to dark molecular gas or H\textsc{i} optical depth corrections). The velocity is given by:
    \begin{equation}
        v_c^2(r) = \Upsilon_{\text{star}} v_{\text{star}}^2(r) + \Upsilon_{\text{gas}} v_{\text{gas}}^2(r)
        \label{eq:bosma_model}
    \end{equation}
    The free parameters are the scaling factors for the stellar and gas components: $\mathbf{p}_{\text{Bosma}} = (\log_{10} \Upsilon_{\text{star}}, \log_{10} \Upsilon_{\text{gas}})$.
\end{enumerate}.

\subsection{Model Comparison Statistics}
\label{subsec:stats}

To compare the relative performance of the best-fit models while accounting for complexity (number of free parameters, $k$), we use the Akaike Information Criterion with a correction for finite sample sizes (AICc). For each model fit, we identify the set of parameters that maximize the posterior probability (equivalent to minimizing $\chi^2$ with flat priors), and calculate the AICc using the resulting minimum chi-squared, $\chi^2_{\text{min}}$:
\begin{equation}
    \text{AICc} = \chi^2_{\text{min}} + 2k + \frac{2k(k+1)}{N-k-1}
    \label{eq:aicc}
\end{equation}
The model with the lowest AICc value is considered statistically preferred for a given galaxy. A difference in AICc of $\Delta\text{AICc} > 10$ between two models is typically considered very strong evidence in favor of the model with the lower value \citep{Burnham2002}.

\section{Results}
\label{sec:results}

Our analysis was conducted on the complete sample of 175 galaxies from the SPARC database. For each galaxy, we fitted all five models to enable a direct and robust comparison.

\subsection{A Decisive Statistical Preference for the Empirical (k=1) Model}
\label{subsec:quant_results}

\begin{figure*}[htbp]
    \centering
    \includegraphics[width=\textwidth]{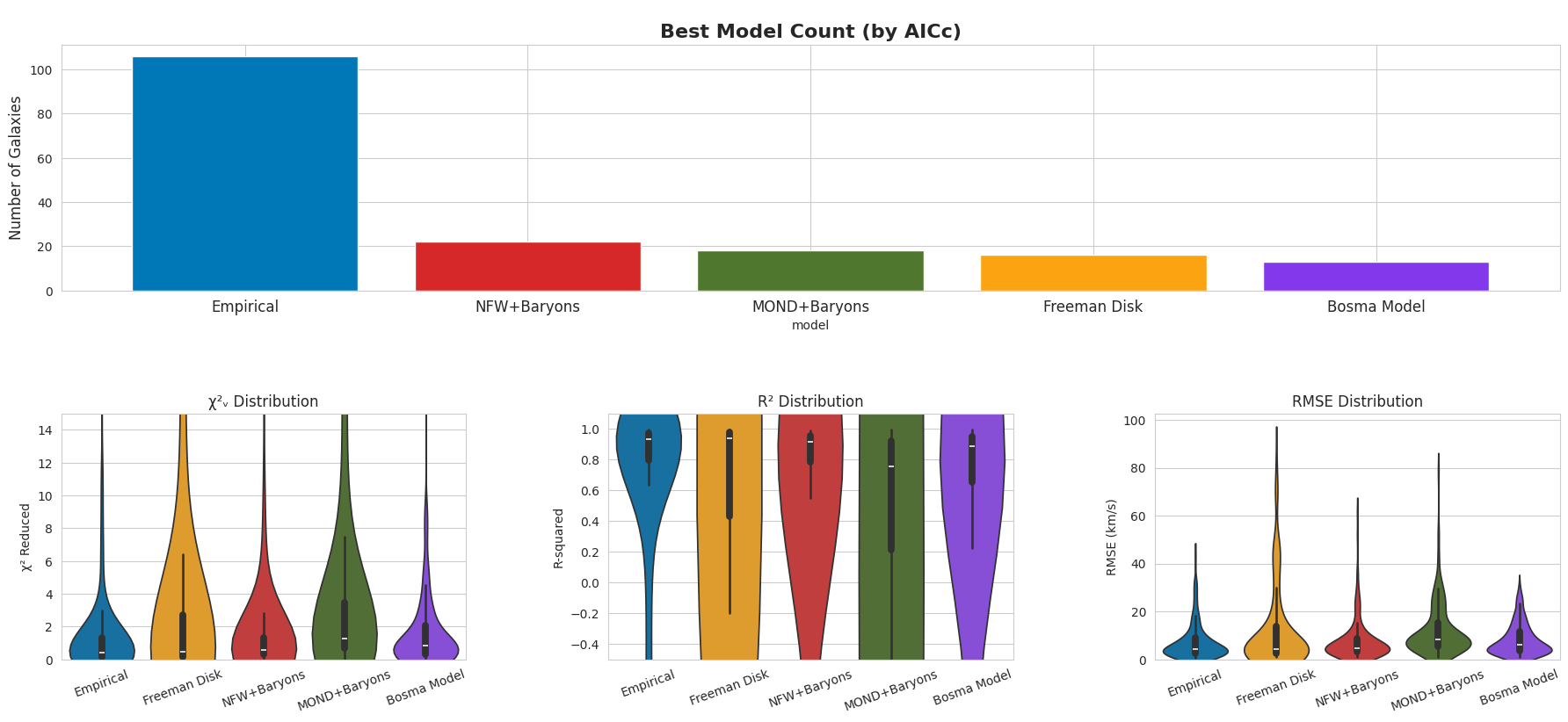} 
    \caption{A visual summary of model performance across the SPARC sample. 
    \textbf{Top panel:} The frequency with which each model was selected as the best description by the AICc, confirming the dominance of the Empirical (k=1) model. 
    \textbf{Bottom panels:} Violin plots showing the distributions of key goodness-of-fit statistics. From left to right, these are the reduced chi-squared ($\chi^2_{\nu}$), the explained variance ($R^2$), and the Root Mean Square Error (RMSE). The compact and low-centered distributions for the Empirical model in the $\chi^2_{\nu}$ and RMSE plots visually underscore its superior statistical performance.}
    \label{fig:summary_performance}
\end{figure*}

The central result of our investigation is a clear, quantitative hierarchy among the tested models. For model selection, we employ the AICc, which rigorously balances a model's goodness-of-fit with its parametric complexity. The top panel of Figure~\ref{fig:summary_performance} presents the primary outcome: the frequency with which each model was selected as the statistically optimal description. The one-parameter Empirical model is decisively preferred, being the winning model for 106 of the 175 galaxies (60.6\%). 

The standard three-parameter NFW model is a distant second, preferred in only 22 cases (12.6\%). This result demonstrates that the additional parametric freedom of the NFW framework is statistically justified for only a small minority of galaxies. The MOND and Freeman Disk models are selected even less frequently, winning in only 18 (10.3\%) and 16 (9.1\%) cases, respectively. Notably, the Bosma (Dark Matter Disk) model, despite linking dynamics directly to the observed gas distribution, is the statistically preferred description for only 13 galaxies (7.4\%) in the sample.

\subsection{Superior Goodness-of-Fit and the Origin of the Preference}
\label{subsec:gof_results}

The statistical preference for the Empirical model is not merely a consequence of its simplicity. Critically, it also provides the most accurate description of the data on average within our comparative framework. Table~\ref{tab:median_stats} reports the median goodness-of-fit statistics for all models tested in this study. The Empirical model achieves a median reduced chi-squared of 0.41. A value substantially below unity signifies that the model's predictions fit the data points more closely than the data's own stated uncertainties. This confirms that the Empirical model is not just the most parsimonious, but also the most accurate overall among the foundational models.

The Bosma model, while physically motivated, yields a median reduced chi-squared of 0.84 nearly double that of the Empirical model. This suggests that strictly coupling the rotation velocity to local irregularities in the gas distribution often degrades the fit compared to the smooth profile of the Empirical law.

For additional context, we include the results from a recent empirical study by \citep{taylor2025empirical}, which used a more flexible, four-parameter polynomial correction factor (DE4-poly) on the same dataset. As expected from its higher parametric freedom, this model achieves a near-perfect median $R^2$, highlighting what is possible with a pure curve-fitting approach. However, our one-parameter Empirical model achieves a highly competitive median $R^2$ of 0.930, demonstrating remarkable descriptive power for its simplicity.

\begin{table}[h!]
\centering
\caption{Model Performance Summary and Statistical Preference. The DE4-poly model from \citep{taylor2025empirical} is included for context, though AICc and $\chi^2_{\nu}$ statistics were not reported in that work.}
\label{tab:median_stats}
\begin{tabular}{l c c c c}
\hline
\hline
Model & Parameters ($k$) & Best Model For (AICc) & Median $\chi^2_{\nu}$ & Median $R^2$ \\
\hline
\textbf{Empirical} & \textbf{1} & {106 (60.6\%)} & \textbf{0.41} & 0.930 \\
NFW+Baryons & 3 & 22 (12.6\%) & 0.59 & 0.915 \\
MOND+Baryons & 1 & 18 (10.3\%) & 1.26 & 0.758 \\
Freeman Disk & 2 & 16 (9.1\%) & 0.45 & 0.937 \\
Bosma Model & 2 & 13 (7.4\%) & 0.84 & 0.885 \\
\hline
DE4-poly \citep{taylor2025empirical} & 4 & \textbf{165 (94.28\%)} & --- & \textbf{0.996} \\
\hline
\end{tabular}
\end{table}

The violin plots in Figure~\ref{fig:summary_performance} visualize these distributions. The plot for the Empirical model is visibly more concentrated at very low $\chi^2_{\nu}$ values, graphically demonstrating its consistent and superior performance. It is noteworthy that the median reduced chi-squared values for the top-performing models are substantially below unity. In a formal statistical sense, this suggests that the total error budget may have been overestimated, likely a consequence of adding a conservative 5.0 km/s systematic error floor to all data points. However, as this error budget was applied uniformly to all five models, it does not impact their relative ranking or the primary conclusion of this work.

\subsection{Qualitative Assessment of Individual Fits}
\label{subsec:qualitative_results}

\begin{figure*}[htbp]
    \centering
    \includegraphics[width=\textwidth]{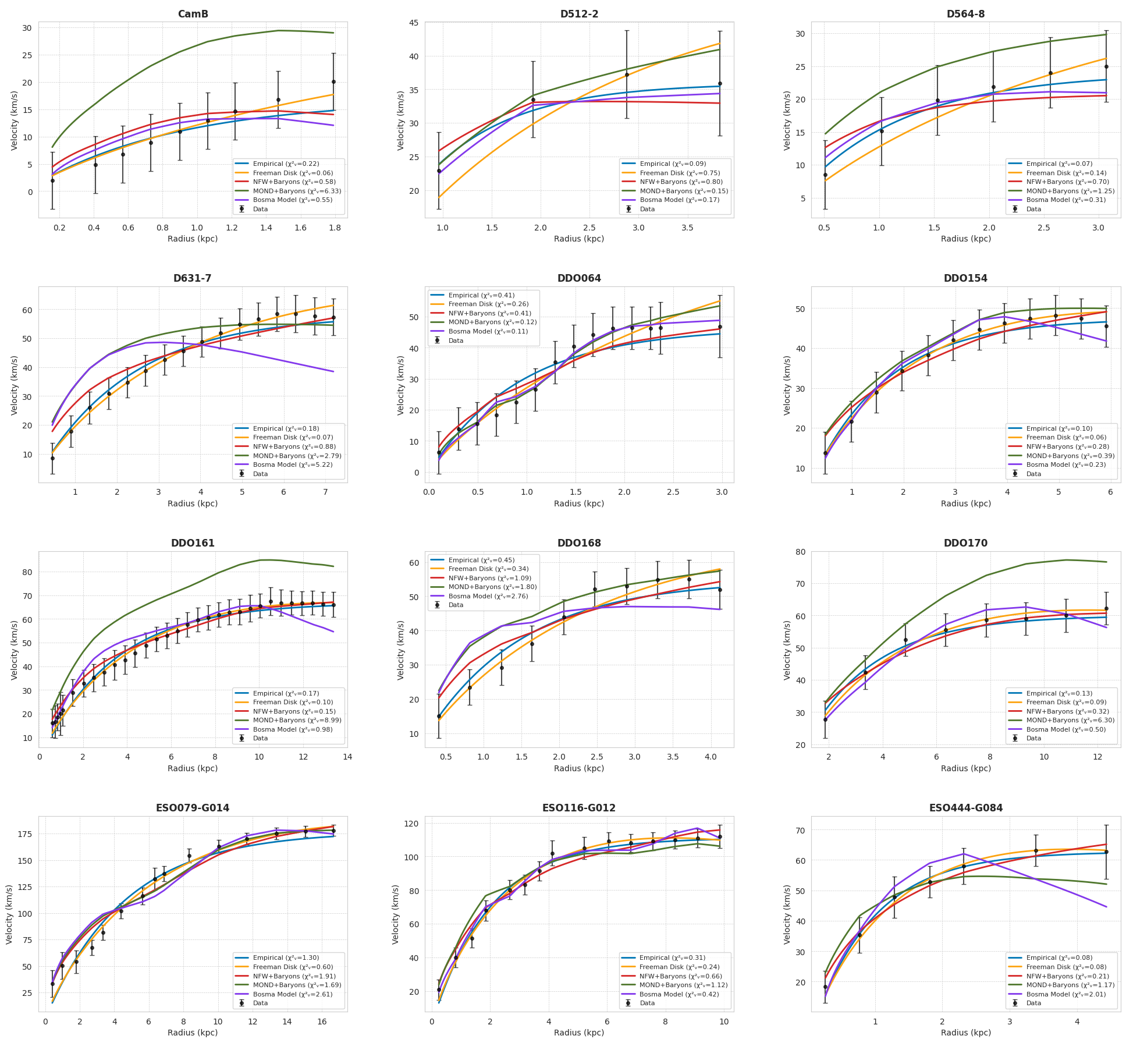}
    \caption{Examples of rotation curve fits for twelve representative SPARC galaxies. Observed data (black points with error bars) are shown alongside the best-fit curves for the four models. The reduced chi-squared ($\chi^2_{\nu}$) values, where available in the legends, confirm the visual impression that the physically-constrained Empirical (k=1) model consistently provides an excellent fit to the data across a wide range of galaxy types.}
    \label{fig:example_fits}
\end{figure*}

To provide qualitative support for these statistical findings, Figure~\ref{fig:example_fits} presents a representative subset of individual model fits for twelve SPARC galaxies. These plots visually confirm the conclusions drawn from the quantitative analysis. In galaxy after galaxy, the constrained one-parameter Empirical model provides a remarkably accurate trace of the observational data. Even in massive, complex systems, the rigid Empirical model often provides a substantially better fit than its competitors. These examples underscore the primary finding of this work: the simple, physically-anchored exponential law is a powerful and highly effective descriptor of observed galaxy kinematics across a vast range of galaxy masses and morphologies.

\section{Discussion}
\label{sec:discussion}

The analyses presented in Section~\ref{sec:results}, based on the complete sample of 175 SPARC galaxies, establish a simple, one-parameter exponential law as a remarkably effective and parsimonious descriptor of galaxy rotation curves. The Empirical model ($k=1$), is physically constrained by the measured flat velocity ($V_{\text{flat}}$) of each galaxy, leaving only a single free parameter ($r_{\text{scale}}$) to describe the entire kinematic profile. Its performance, rigorously evaluated through goodness-of-fit statistics (median $\chi^2_{\nu} = 0.41$) and information-theoretic criteria (AICc), demonstrates a clear superiority against established benchmark models. These benchmarks include the standard three-parameter NFW profile (median $\chi^2_{\nu} = 0.59$), the one-parameter MOND formulation (median $\chi^2_{\nu} = 1.26$), the classical two-parameter Freeman Disk model (median $\chi^2_{\nu} = 0.45$), and the two-parameter Bosma model (median $\chi^2_{\nu} = 0.84$). The profound success of the Empirical law, particularly its selection as the best model by AICc for a decisive majority (60.6\%) of the sample, highlights its significance. This section delves into the interpretation of these findings and contextualizes their importance within galactic dynamics.

The results of this study can be contextualized by comparison to other recent empirical investigations. For instance, the `detection factor' proposed by \citep{taylor2025empirical}, a flexible, four-parameter polynomial correction achieves an exceptionally high median $R^2 \approx 0.996$ on the SPARC dataset. While this demonstrates a tight fit, it is characteristic of the model's higher degree of parametric freedom. In contrast, the one-parameter law presented in this work provides a far more parsimonious description of the data. The superior performance of our model under the AICc framework, which explicitly penalizes complexity, suggests that our single parameter, $r_{\text{scale}}$, captures the essential dynamics without the risk of overfitting inherent in more flexible models. This reinforces our primary conclusion that a remarkably simple, physically-constrained law provides the most statistically efficient description of galactic rotation curves.

\subsection{The Power of Parsimony: Efficiency as the Deciding Factor}
\label{subsec:discuss_empirical}

A striking outcome of our analysis is that while multiple models can provide excellent fits to the data, the principle of parsimony decisively favors the simplest one. The Empirical model's median reduced chi-squared of 0.41 is the best among all competitors, indicating that its rigid, one-parameter form provides the closest overall fit to the observational data. This is a non-trivial result; it demonstrates that a simple mathematical law, when anchored to a real physical quantity ($V_{\text{flat}}$), has more descriptive power than more complex models with greater freedom.

The comparison with the NFW+Baryons ($k=3$), Freeman Disk ($k=2$), and Bosma ($k=2$) models is particularly illuminating. While these frameworks capture the general kinematic trends, our results show they are statistically preferred in only 22 (12.6\%), 16 (9.1\%), and 13 (7.4\%) cases, respectively. This implies that the additional flexibility afforded by extra free parameters—or the strict coupling to gas density in the Bosma case—is often not statistically justified by the data. In a majority of cases, the one-parameter Empirical law provides an explanation that is not only of equal or superior quality but is also vastly more efficient.

Furthermore, the head-to-head comparison between the two one-parameter models, the Empirical law and MOND, provides a clear verdict. With equal complexity ($k=1$), goodness-of-fit becomes the sole arbiter of performance. The Empirical model's median $\chi^2_{\nu}$ of 0.41 is vastly superior to that of MOND (1.26). This demonstrates that, for the SPARC sample, the exponential form of the Empirical law is a fundamentally better mathematical description of galaxy kinematics than the interpolating function used in MOND.

\subsection{Systematic Failures and Model Robustness}
\label{subsec:discuss_failures_discussion}

While the Empirical law demonstrates remarkable descriptive efficacy, a complete analysis requires a transparent examination of its limitations. To rigorously assess robustness, we analyzed the incidence of catastrophic model failures, defined as cases where a model yielded a coefficient of determination $R^2 < 0$ or failed to converge. The standard NFW halo proved the most robust against total failure (19 galaxies; 10.8\%), followed closely by the Bosma model (21 galaxies; 12.0\%) and the Empirical law (26 galaxies; 14.8\%). In contrast, the MOND (36 failures) and Freeman Disk (39 failures) models struggled significantly more often. 

A systematic review of these failures reveals a high degree of overlap across all frameworks, pointing to a distinct physical regime where simple axisymmetric models break down. The most dramatic failures consistently occur in massive, high-surface-brightness spiral galaxies. Specifically, systems such as \textbf{NGC 0891, NGC 5005, NGC 5033, NGC 5371, and NGC 5907} appear as common failures across the Empirical, Bosma, and NFW models. Furthermore, the Empirical model specifically struggled with complex systems like \textbf{NGC 2841, NGC 6946, and NGC 7331}. 

Quantitatively, the poorest fits are found in the most rapidly rotating galaxies ($V_{\text{flat}} \ge 200$ km s$^{-1}$). For instance, \textbf{UGC 02487} ($V_{\text{flat}} \approx 333$ km s$^{-1}$), \textbf{UGC 09133} ($V_{\text{flat}} \approx 229$ km s$^{-1}$), and \textbf{NGC 6674} ($V_{\text{flat}} \approx 242$ km s$^{-1}$) are among the worst-fit objects for the NFW, Freeman, and MOND models alike. These massive systems likely possess complex central dynamics, including significant bulges, bars, or strong non-circular motions, which are not adequately captured by any of the simple, axisymmetric models tested. This finding delineates a physical boundary for the applicability of these frameworks and underscores that the dynamics of the most massive disk galaxies represent a distinct challenge for current modeling paradigms.

In light of these limitations, it is crucial to acknowledge that Equation 1 is, at this stage, a phenomenological law rather than a derived solution from a fundamental Lagrangian. Unlike the NFW profile, which emerges from cosmological N-body simulations of cold dark matter \cite{Navarro1997}, or the Freeman disk, which is derived from Newtonian potential theory \cite{freeman1970disks}, our model was established through rigorous statistical experimentation on the SPARC database. Initially exploring a broader parametric space, our preliminary analysis uncovered a robust correlation between the model's fitted asymptotic velocity and the independently measured flat velocity ($V_{\text{flat}}$) inherent in the dataset. This discovery allowed us to evolve the equation into a \textbf{hybrid formulation}: rather than treating the asymptotic velocity as a free parameter to be minimized, we integrate the observed $V_{\text{flat}}$ directly as a fixed physical constraint. This reduction in dimensionality not only enhances statistical parsimony but also anchors the model in observational reality. The equation's persistent statistical superiority over traditional physically derived models suggests that it captures a fundamental kinematic reality specifically, the exponential rise to stability that current theories do not fully encapsulate.

\subsection{The Disk-Based Paradigm and the Bosma Effect}
Our results lend strong statistical support to the view that the dynamics of disk galaxies are governed by mechanisms confined to the galactic plane, rather than an extended spherical halo. This aligns fundamentally with recent investigations into the ``Dark Matter Disk'' (DMD) and the ``Bosma effect,'' which postulate that the rotation curve shape is dictated by the baryonic mass distribution within the disk itself \cite{labini2025exploring, SylosLabini2024, Bosma1981}. By demonstrating that a model physically constrained by the disk's flat velocity ($V_{\text{flat}}$) outperforms halo-based models, our work reinforces the core premise of these investigations: the visible matter might be the primary driver of galactic kinematics, likely orchestrated by an unknown characteristic of the fabric geometry that enforces global stability.

To test the specific formulation of this paradigm, we explicitly modeled the rotation velocity as a linear combination of the stellar and gas contributions ($v_c^2 = \Upsilon_s v_{\text{stars}}^2 + \Upsilon_g v_{\text{gas}}^2$). Crucially, we adopted a \textbf{generalized formulation} ($k=2$) that allows both the stellar and gas mass-to-light ratios to vary freely. This approach grants the Bosma model maximal flexibility to account for unmodeled factors, such as molecular gas scaling or H\textsc{i} optical depth corrections. Despite this parametric advantage, our AICc analysis reveals it is the statistically preferred description for only 7.4\% of the sample (13 galaxies). In comparison, the standard NFW halo model performs relatively better (12.6\% preference), largely due to its even higher parametric freedom ($k=3$) which allows it to accommodate kinematic features that the strictly gas-coupled Bosma model cannot.

However, the most significant finding is that the single-parameter Empirical Law ($k=1$) decisively outperforms both. Preferred for 60.6\% of the sample with the lowest median chi-squared ($\chi_{\nu}^2 = 0.41$), it demonstrates that neither the flexibility of the halo model nor the strict gas-tracing of the Bosma model is necessary to describe galactic rotation. We interpret this as evidence that while the disk's potential dominates, the dynamics are not slavishly coupled to stochastic irregularities in the local gas distribution. Instead, the success of our monotonic exponential function implies that the underlying fabric geometry may act as a smoothing substrate, filtering local baryonic fluctuations to produce a coherent, stable velocity profile.

\subsection{Broader Context and Future Prospects}
\label{subsec:discuss_future}

This study positions the one-parameter law $v(r) = V_{\text{flat}}(1 - \exp(-r / r_{\text{scale}}))$ as a fundamental descriptor of disk galaxy dynamics. The history of physics is replete with examples of empirically discovered laws—such as Kepler's laws or the Rydberg formula that preceded and guided the development of deeper physical theories. In this context, the Empirical law serves as a powerful and challenging new benchmark. Any successful theory of galaxy formation or modified gravity must not only explain the existence of flat rotation curves, but it must also explain why their kinematic profiles so closely and consistently follow this simple, one-parameter exponential form.

Future research should build upon this discovery:
\begin{enumerate}
    \item \textbf{Establishing Scaling Relations:} A detailed analysis of the correlations between the best-fit parameter, $r_{\text{scale}}$, and independently measured global galaxy properties (e.g., baryonic mass, luminosity, size) is now a critical next step.
    \item \textbf{Exploring Physical Underpinnings:} The consistent success of this simple mathematical form strongly motivates theoretical investigation into its physical origins.
    \item \textbf{Application as a Standardized Tool:} Given its accuracy and simplicity, the Empirical law can serve as a new standard method to parameterize observed rotation curves, facilitating robust comparisons between different surveys and simulations.
\end{enumerate}

\section{Conclusion}
\label{sec:conclusion}

In this investigation, we have rigorously assessed the applicability of a concise, one-parameter empirical law for describing the rotation curves of spiral galaxies. Through a detailed Bayesian MCMC analysis of the complete SPARC database (175 galaxies), we have demonstrated that this constrained Empirical model exhibits superior statistical efficiency and accuracy when compared against several established theoretical frameworks, including the Dark Matter Disk (Bosma) model.

Our principal findings, derived from this comprehensive analysis, can be summarized as follows:

\begin{enumerate}
    \item The one-parameter Empirical model is the statistically preferred model for the majority of the SPARC sample. Model selection based on the Akaike Information Criterion (AICc) identifies it as the optimal description for a decisive 106 galaxies (60.6\%). This significantly outperforms the NFW+Baryons model (22 galaxies; 12.6\%), the MOND model (18 galaxies; 10.3\%), and the Freeman Disk model (16 galaxies; 9.1\%). Notably, the Bosma model is preferred in only 13 cases (7.4\%), indicating that a direct coupling between gas density and rotation velocity is rarely the statistically optimal description.

    \item The Empirical model provides the most accurate overall fit to the data. Quantitative evaluation reveals that it achieves the best concordance with the observations, evidenced by a median reduced chi-squared ($\chi^2_{\nu}$) of 0.41. This value is superior to those obtained for a Freeman Disk (0.45), a standard Navarro-Frenk-White profile (0.59), the Bosma model (0.84), and a MOND formulation (1.26).

    \item The model's single free parameter, $r_{\text{scale}}$, is well-constrained and physically meaningful. The distribution of best-fit $r_{\text{scale}}$ values across the sample forms a well-behaved, unimodal distribution that peaks at 1--2 kpc, providing a direct empirical measurement of the characteristic transition scale of galaxy rotation curves.
\end{enumerate}

The central contribution of this work is the demonstration that a simple phenomenological law, when physically anchored to real data, can outperform more complex models in statistical efficiency. Our analysis also delineates a clear boundary for its applicability, as all tested models systematically fail to describe the dynamics of the most massive, rapidly-rotating galaxies. Irrespective of any deeper physical interpretation, this work establishes the constrained exponential function as a highly effective and statistically superior benchmark for the astronomical community.

\section*{Ethics Declaration}
Not applicable.
\section*{Funding}
This research received no specific grant from any funding agency in the public, commercial, or not-for-profit sectors.
\bibliographystyle{plainnat}
\bibliography{ref_4}
\end{document}